\documentclass[%
aps,
reprint,
prb,
amsfonts,
amssymb,
amsmath
]{revtex4-2}

\usepackage{bm}

\usepackage{graphicx}
\usepackage[%
colorlinks=true,
allcolors=blue,
]{hyperref}

\newcommand{\ndag}{{\vphantom{\dag}}}
\DeclareMathOperator{\Tr}{Tr}

\begin{document}

\title{Intersublattice entanglement entropy of ferrimagnetic spin chains}
\date{\today}
\author{Jongmin Y. \surname{Lee}}
\affiliation{Department of Physics, Korea Advanced Institute of Science and Technology, Daejeon 34141, Korea}
\author{Se Kwon \surname{Kim}}
\email{sekwonkim@kaist.ac.kr}
\affiliation{Department of Physics, Korea Advanced Institute of Science and Technology, Daejeon 34141, Korea}

\begin{abstract}
	Ferrimagnets are antiparallel-ordered magnetic states in a bipartite lattice with two alternating unequal spins, which exhibit both ferromagnetic and antiferromagnetic properties.
	Several theoretical studies have explored the magnetic properties of ferrimagnets, but the entanglement entropy of ferrimagnets with arbitrary spin combinations has not been studied.
	In this study, we analytically derive the intersublattice entanglement entropy of a ferrimagnetic spin chain using the method that has been applied to the antiferromagnetic case.
	The analytical results are numerically verified using the density matrix renormalization group.
	Going beyond the results for antiferromagnets, the entanglement entropy of ferrimagnets for fixed anisotropy is shown to solely depend on the difference of spins relative to its geometric average, and the quantity is shown to be stable against small parameter variations.
\end{abstract}

\maketitle

\section{Introduction}
\label{sec:1}

Ferrimagnets are antiparallel-aligned magnetic states in a bipartite lattice constituted by two sublattices of unequal spins~\cite{kim_ferrimagnetic_2022}. Owing to the unique nature of ferrimagnets originating from the unbalanced spins in opposite direction that they exhibit both antiferromagnetic and ferromagnetic properties, several studies on their characteristics and applications have been conducted.
From the perspective of many-body physics, thermodynamic quantities~\cite{pati_low-lying_1997,niggemann_mixed_1997,*niggemann_mixed_1998,maisinger_thermodynamics_1998,fukushima_thermodynamic_2004,wu_exact_2011} and phases~\cite{alcaraz_critical_1997,solano-carrillo_entanglement_2011} of mixed spin chains have been theoretically studied.
On a different side, ferrimagnets are being actively studied in spintronics as material platform for spin devices~\cite{zhang_ferrimagnets_2023}.
For such applications, achieving theoretical understanding of ferrimagnetic states in comparison to antiferromagnetism is important~\cite{baltz_antiferromagnetic_2018}.

Recently, quantum informational approaches in condensed matter physics has been found to be useful~\cite{zeng_quantum_2019}.
For example, information-theoretic quantities can be utilized to analyze the characteristics of low energy states of quantum many-body systems~\cite{amico_entanglement_2008}.
One of these quantities is entanglement entropy, which measures the degree of non-separability of density operators over a bipartite system.
Entanglement entropy can be used to characterize a quantum phase transition~\cite{osborne_entanglement_2002,osterloh_scaling_2002} or to detect a topological order~\cite{kitaev_topological_2006,levin_detecting_2006} in the ground state of a many-body system.
Thanks to the development of tensor networks, a category of numerical methods that can efficiently simulate highly entangled many-body states using tensors, one can extract physical and informational quantities including entanglement entropy from the tensor networks states~\cite{orus_tensor_2019}.
Consequently, analytic approaches to the ground states of many-body systems can be tested numerically.

\begin{figure}[t]
	\includegraphics[width=0.48\textwidth]{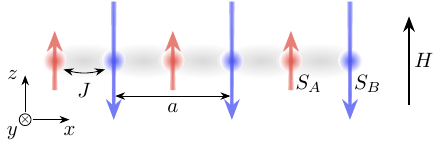}
	\caption{%
		\label{fig:1}
		Schematic illustration of a ferrimagnetic spin chain for $N = 3$ and $S_B / S_A = 2$ case.
		Each sublattice $A$ (red) and $B$ (blue) consists of spin-$S_A$ and spin-$S_B$ particles respectively, and the lattice spacing is equal to $a$.
		Heisenberg interactions are imposed between nearest-neighbor sites, and an external magnetic field is applied along the $z$-axis.
	}
\end{figure}

In Ref.~\cite{hartmann_intersublattice_2021}, entanglement entropy between sublattices, referred to as intersublattice entanglement entropy, was analytically studied in an antiferromagnet model with uniaxial anisotropy.
They derived an analytic expression of the intersublattice entanglement entropy for arbitrary dimensions and numerically verified it in the 1D case using density matrix renormalization group (DMRG), one of tensor network algorithms~\cite{white_density_1992,white_density-matrix_1993,schollwock_density-matrix_2011}.
The study discusses that the analytic relation approximates the entanglement entropy with high precision for spins $S \ge 3/2$ in the large anisotropy region.
Mathematically, Lieb-Schultz-Mattis theorem explains the degeneracy of ground states in antiferromagnets as spin varies~\cite{lieb_two_1961,affleck_proof_1986,tasaki_physics_2020}, but exact solutions or the entanglement structure of ferrimagnets are mainly studied for $(1/2,S)$ combinations~\cite{wang_thermal_2006,hao_entanglement_2007,guo_entanglement_2011,vargova_unconventional_2021,vargova_effect_2022}.
To investigate the difference between antiferromagnet spin chains and ferrimagnet spin chains for all possible spin combinations, we extended the method presented in Ref.~\cite{hartmann_intersublattice_2021} into a ferrimagnetic spin chain, which is schematically illustrated in Fig.~\ref{fig:1}.

The rest of the paper is organized as follows.
We define a ferrimagnet spin chain model and diagonalize the Hamiltonian using the squeezed magnon approximation in section~\ref{sec:2}.
Using the ground state, we derive an analytic expression for intersublattice entanglement entropy [Eq.~\eqref{eq:entropy}] in section~\ref{sec:3}.
The relation matches the one for antiferromagnets derived in Ref.~\cite{hartmann_intersublattice_2021} if the alternating spin sizes are equal.
Then we compare the analytic relation with numerical data using DMRG in section~\ref{sec:4} to check the accuracy.
We finish the paper with discussions on the characteristics of ferrimagnets and a few open questions in section~\ref{sec:5}.

\section{Model}
\label{sec:2}

We consider a ferrimagnet as a periodic uniaxial mixed spin chain in an external magnetic field along the $z$ axis consisting of $N$ number of spin-$S_A$ ($S_B$) particles on sublattice $A$ ($B$) aligned along the ($-$)$z$ axis, described by the Hamiltonian
\begin{eqnarray}
	\mathcal{H} &=& \frac{J}{\hbar^2} \sum_{i=1}^{N} \left( \bm{S}_{A,i} \cdot \bm{S}_{B,i} + \bm{S}_{B,i} \cdot \bm{S}_{A,i+1} \right) \nonumber\\
	&&- \sum_{ \alpha \in \{A, B\} } \sum_{i=1}^{N} \left( \frac{K}{\hbar^2} \left( S_{\alpha,i}^{(z)} \right)^2 + \gamma_\alpha H S_{\alpha,i}^{(z)} \right),
\end{eqnarray}
where $\bm{S}_{A,N+1} = \bm{S}_{A,1}$, $J>0$ is the exchange coupling, $K>0$ is the easy-axis anisotropy energy, $\gamma_\alpha<0$ is the gyromagnetic ratio, and $H \hat{\mathbf{z}}$ is the external magnetic field.
Figure~\ref{fig:1} represents the model for the case where $N=3$ and $S_B / S_A = 2$.
Hereafter, we assume $S_A \le S_B$ without loss of generality and both gyromagnetic ratios are equal to a constant value $\gamma$.

We apply the Holstein--Primakoff transformation~\cite{holstein_field_1940} to express the Hamiltonian in quadratic form (up to a constant) in the local bosonic operators $a_i$ ($b_i$) at the $i$th site in the sublattice $A$ ($B$) that satisfy canonical commutation relations, under large spin assumption.
The quadratic form can be reformulated into a block diagonal form using Fourier transformation, where $N$ independent blocks are written in terms of bosonic operators $a_{k} = N^{-1/2} \sum_{j=1}^{N} e^{-ikja} a_j$ and $b_{-k}^\dag = N^{-1/2} \sum_{j=1}^{N} e^{-ikja} b_j^\dag$ over each sublattice, which are labeled by a wavenumber $k \in \left\{ -\pi/a , \ldots, \pi(N-2)/Na \right\}$.
Then each block can be diagonalized using the Bogoliubov transformation~\cite{bogoljubov_new_1958} and the Hamiltonian is rewritten as
\begin{equation}
	\mathcal{H} = \sum_{k} \left( E_k^{-} \alpha_k^\dag \alpha_k^\ndag + E_k^{+} \beta_k^\dag \beta_k^\ndag \right) + E_{\textrm{GS}},
\end{equation}
where $\alpha_k^{(\dag)}$ and $\beta_k^{(\dag)}$ are bosonic operators which satisfy canonical relations.
The magnon gaps $E_k^\pm$ and the ground state energy $E_{\textrm{GS}}$ are given as
\begin{eqnarray}
	E_k^\pm &=& E_k^0 \pm B, \quad E_k^0 = \sqrt{A^2 - C_k^2}, \\
	\frac{E_{\mathrm{GS}}}{N} &=& -2JS_A S_B - K\left( S_A^2 + S_B^2 \right) - \gamma \hbar H(S_A - S_B) \nonumber\\
	&&- \frac{1}{N}\sum_k \left( A - \sqrt{ A^2 - C_k^2 } \right),
\end{eqnarray}
for $A = (S_A + S_B)(J + K)$, $B = (S_A - S_B)(J - K) - \gamma \hbar H$, and $C_k = 2J\sqrt{S_A S_B} \cos(ka/2)$.
To guarantee the magnon gaps $E_k^\pm$ are positive for all $k$, we restrict the range of external magnetic field as $0 \le H \le 2K(S_B - S_A)/|\gamma|\hbar$.

The bosonic operators $\alpha_k$ and $\beta_k$ are called squeezed magnon operators since they are related to $a_k^\ndag$ and $b_{-k}^\dag$ via $\begin{pmatrix}
	\alpha_k^\ndag & \beta_k^\dag
\end{pmatrix} = S(\phi_k) \begin{pmatrix}
	a_k^\ndag & b_{-k}^\dag
\end{pmatrix} S^\dag(\phi_k)$ using the two-mode squeezing operator $S(\phi_k)=e^{\left( a_k^\ndag b_{-k}^\ndag - a_k^\dag b_{-k}^\dag \right)\phi_k}$, where a squeezing parameter $\phi_k \ge 0$ is determined by $\cosh\phi_k = (C_k/A)^{-1}$~\cite{yun-xia_entanglement_2008}.
For coherent states, squeezing changes the ratio between uncertainties of the quadratures $\left( d_k^\ndag + d_k^\dag \right) / \sqrt{2}$ and $\left( d_k^\ndag - d_k^\dag \right) / \sqrt{2}i$ for $d_k = \left( a_k + b_{-k} \right) / 2$, while preserving their product.
Since $C_k/A = 1$ holds only for antiferromagnets without anisotropy ($S_A = S_B$ and $K=0$) with $k=0$, the squeezing parameter does not vanish for ferrimagnets.

To represent a state of the system, one can consider two magnon number basis $| m, n \rangle_{\textrm{sub},k}$ and $| m, n \rangle_{\textrm{sq},k}$ that corresponds to the ordered pairs of bosonic operators $(a_k, b_k)$ and $(\alpha_k, \beta_k)$ for each $k$, respectively.
Then the ground state of the Hamiltonian becomes $| \textrm{GS} \rangle = \bigotimes_{k} | 0, 0 \rangle_{\textrm{sq},k}$, which is a product of two-mode squeezed vacuum states for each $k$.
Using the relation $| 0,0 \rangle_{\textrm{sq},k} = S(\phi_k) | 0,0 \rangle_{\textrm{sub},k}$ in Ref.~\cite{caves_new_1985,*schumaker_new_1985}, the ground state can be expressed in terms of a sublattice Fourier number basis for each $k$ as \begin{equation}
	| \textrm{GS} \rangle = \bigotimes_{k} \frac{1}{\cosh\phi_k} \sum_{n=0}^{\infty} (-\tanh\phi_k)^n | n,n \rangle_{\textrm{sub},k}.
\end{equation}

The ground state energy $E_{\textrm{GS}}$ is analytically expressed under the thermodynamic limit $N \to \infty$.
An analytic expression for the energy of a ferrimagnet spin chain per unit cell divided by $JS_A S_B$, is given as follows:
\begin{eqnarray}
	\frac{E_{\textrm{GS}}}{JN\sigma^2} &\approx& -\frac{K}{J}\left(\frac{\Delta}{\sigma}\right)^2 - 2\left(1 + \frac{K}{J}\right) \nonumber\\
	&&\!\!\!+ \frac{1}{\sigma} \left( \frac{\gamma\hbar H}{J}\left(\frac{\Delta}{\sigma}\right) - \frac{1}{2m}\left( 1 - \frac{2}{\pi} E(m) \right) \right)\!,\, \label{eq:energy}
\end{eqnarray}
where the approximation symbol represents thermodynamic limit, $\Delta = S_B - S_A$ is a spin difference, $\sigma = \sqrt{S_A S_B}$ is a geometric mean of spins, $m^{-1} = (1 + K/J)\sqrt{1 + (\Delta/2\sigma)^2}$ and $E(m) = \int_0^{\pi/2} d\theta\,\sqrt{1 - m^2 \sin^2 \theta}$ is a complete elliptic integral of the second kind.
The right-hand side of Eq.~\eqref{eq:energy} is a function of anisotropy $K/J$ and a scaled spin difference $\Delta/\sigma$, up to $O(\sigma^{-1})$ correction terms.
When the two spins are the same, i.e., $S_A = S_B$ with $\Delta/\sigma = 0$, the expression matches the result of antiferromagnets~\cite{hartmann_intersublattice_2021}.
The first term demonstrates the ferrimagnetic nature through its dependency on the scaled spin difference.
Under the assumption of zero external magnetic field, the correction term can be further ignored when $\sigma \gg 1$ or $K/J \gg 1$.

\section{Entanglement Entropy}
\label{sec:3}

Entanglement entropy is a measure of quantum information for a bipartite quantum system that quantifies the quantum entanglement between two subsystems.
For a state described by a density operator over the composite system, entanglement entropy is defined as the von Neumann entropy of the reduced density operator for one subsystem~\cite{zeng_quantum_2019}.
We obtain the reduced density matrix of the ground state by partial tracing out the subsystem $B$ to calculate the entanglement entropy between the sublattices $A$ and $B$, or an intersublattice entanglement entropy of the ferrimagnetic spin chain,
\begin{eqnarray}
	\rho_A &=& \Tr_B |\textrm{GS}\rangle\langle\textrm{GS}| \nonumber\\
	&=& \bigotimes_k \left( \sum_{n\ge0} \frac{\tanh^{2n}\phi_k}{\cosh^2\phi_k} | n \rangle_{A,k}\langle n | \right),
\end{eqnarray}
where the projection operator $| n\rangle_{A,k}\langle n |$ is in the Fourier number basis over sublattice $A$ for each $k$.
The entanglement entropy is given by
\begin{equation}
	S_{\textrm{EE}} = -\Tr_A \rho_A \ln\rho_A = \sum_k s_k,
\end{equation}
where $s_k = \cosh^2\phi_k \ln\left(\cosh^2\phi_k\right) - \sinh^2\phi_k \ln\left( \sinh^2\phi_k \right)$.

Our main result, an analytic expression for the intersublattice entanglement entropy per unit cell, is derived in the thermodynamic limit:
\begin{equation}
	\frac{S_{\textrm{EE}}}{N} \approx \frac{1}{\kappa} \int_0^{\kappa} dk\, \left( u_k^{+} \ln u_k^{+} - u_k^{-} \ln u_k^{-} \right), \label{eq:entropy}
\end{equation}
where the approximation symbol indicates the thermodynamic limit, $\kappa = \pi\sqrt{m}/2\sqrt{2}$, and $u_k^{\pm} = \left( \left( 1 - \left( m - k^2 \right)^2 \right)^{-1/2} \pm 1 \right)\!\!/2$ with $m$ the inverse of $(1+K/J)\sqrt{1+(\Delta/2\sigma)^2}$.
This indicates that entanglement entropy per unit cell of a ferrimagnetic spin chain is determined by $m$, which is a function of the ratio $K/J$ and the scaled spin difference $\Delta/\sigma = (S_B - S_A) / \sqrt{S_A S_B}$. 
Hence the external magnetic field $H$ cannot affect the entanglement entropy per unit cell, but instead the ferrimagnetic quantity $\Delta/\sigma$ that characterized the imbalance between the two spins is reflected on it.
Note that Eq.~\eqref{eq:entropy} matches the result of antiferromagnetic case when $\Delta/\sigma=0$~\cite{hartmann_intersublattice_2021}, as with Eq.~\eqref{eq:energy}.

\section{Numerical Verification}
\label{sec:4}

Density matrix renormalization group (DMRG) is an efficient tensor network method for investigating low energy physics of 1D gapped systems~\cite{white_density_1992,white_density-matrix_1993,schollwock_density-matrix_2011}.
The state and Hamiltonian are represented as a matrix product state (MPS) and a matrix product operator (MPO).
DMRG iterates two-site update with truncating the matrix up to the bond dimension, resulting in a polynomial time complexity for search of ground states. At the end of the algorithm, we obtain the ground state and its energy for the given Hamiltonian.
In many cases, the MPS is chosen as a canonical form to fix the gauge of the MPS that satisfies isometry conditions.
A mixed canonical form, which includes the Schmidt coefficients at the cut between the left and right subsystems, is useful for extracting the entanglement entropy since Schmidt decomposition of a state over a bipartite system allows us to calculate the von Neumann entropy.

\begin{figure}
	\includegraphics[width=0.48\textwidth]{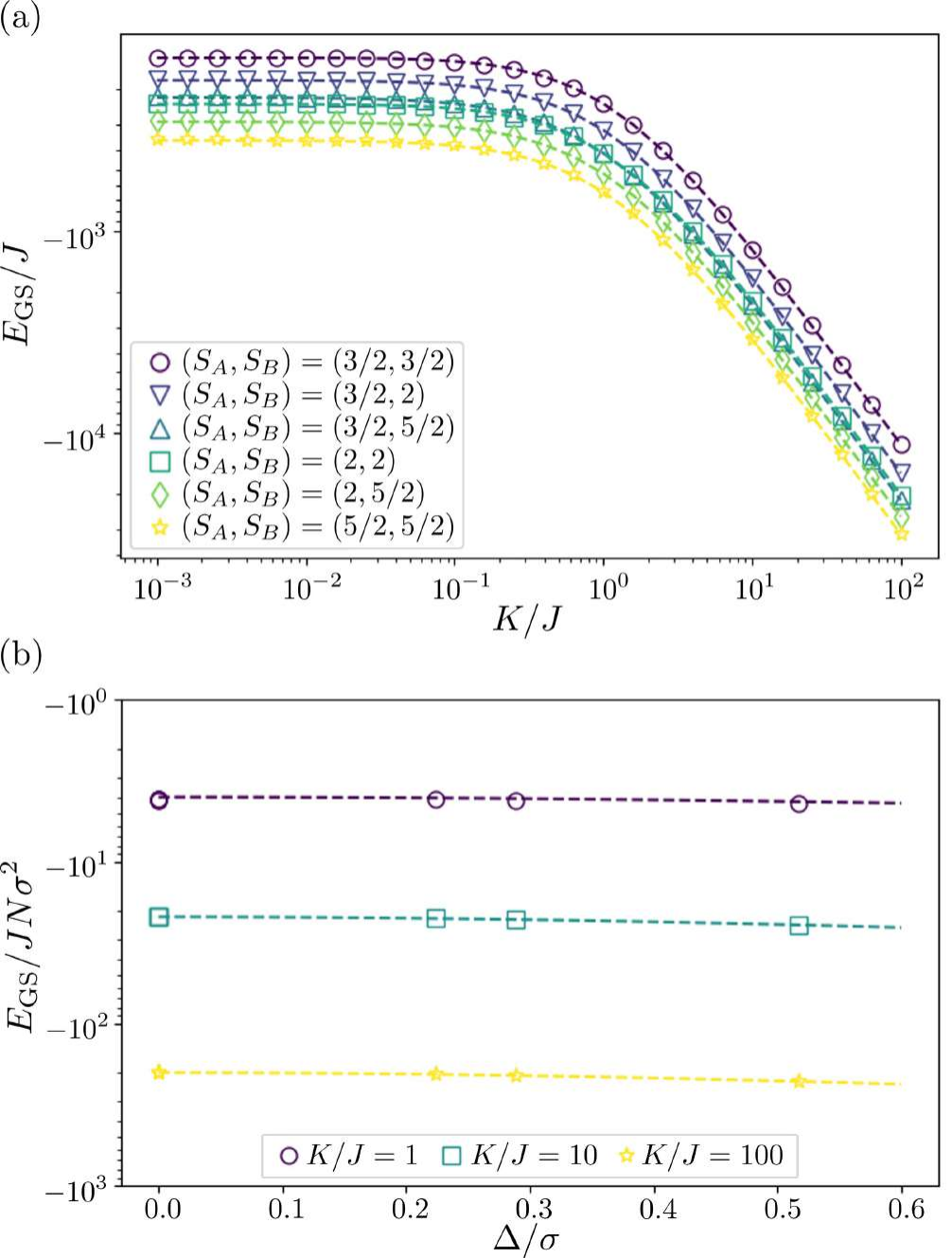}
	\caption{%
		\label{fig:2}
		Comparison of ground state energy between the analytic approximation [Eq.~(\ref{eq:energy})] (dotted line) and DMRG simulation (marker) with $50$ sites considering six possible combinations of spins from $3/2$ to $5/2$.
		(a) Plot of $E_{\textrm{GS}}/J$ as a function of the ratio $K/J$ for each combination of spins.
		(b) Plot of $E_{\textrm{GS}}/JN\sigma^2$, a ground state energy per unit cell divided by $J\sigma^2$ as a function of the ratio $\Delta/\sigma = (S_B - S_A) / \sqrt{S_A S_B}$, for $K/J = 1, 10, 100$.
		The dotted lines decline as $\Delta/\sigma$ increases.
	}
\end{figure}

In this research, DMRG is implemented using Python package `ncon'~\cite{pfeifer_ncon_2015} and library `NumPy'~\cite{harris_array_2020}.
We considered antiferromagnetic and ferrimagnetic spin chains with $J=1$ and restricted external magnetic field situations, consisting of $N=25$ unit cells where the spin size ranges from $3/2$ to $5/2$.
Range of the interaction strength is chosen as $10^{-3} < K/J < 10^2$.
Initial MPS for the case of $K/J=10^{-3}$ is calculated with iterative diagonalization under the assumption of a large magnetic field $|\gamma|\hbar H/J = 10^4$.
Then we repeated DMRG on that state by decreasing the exponent of magnetic field until $|\gamma| \hbar H/J = 10^{-5}$ to ensure numerical stability.
As we increase the interaction strength, the initial MPS for applying DMRG algorithm is set to the ground state corresponding to the preceding $K/J$ value.

Figure~\ref{fig:2} compares the ground state energy of the ferrimagnet spin chain between the analytic relation~\eqref{eq:energy} and DMRG calculation to $K/J$ and $\Delta/\sigma$.
In Fig.~\ref{fig:2}(a), we plotted $E_{\textrm{GS}}$ against $K/J$ for each spin configuration, where the shape of graph changes from a plateau to a linearly decreasing line as $K/J$ increases.
Near $K/J=1$, the graphs for configurations $(S_A, S_B) = (3/2,5/2), (2,2)$ intersect each other.
This implies that the energy correction term in an antiferromagnetic case is stronger than that of the ferrimagnet when $K/J<1$, and the quadratic term in $\Delta/\sigma$ becomes a leading term when $K/J>1$.
In Fig.~\ref{fig:2}(b), $E_{\textrm{GS}}/JN\sigma^2$ is plotted against the scaled spin difference $\Delta/\sigma$.
At $\Delta/\sigma=0$, three markers correspond to the antiferromagnetic spin chains with spins $3/2$, $2$, $5/2$ overlap.
The dotted line is drawn without the energy correction term, which implies that we can ignore the term $O(\sigma^{-1})$ for $K/J>1$.

\begin{figure}[t]
	\includegraphics[width=0.48\textwidth]{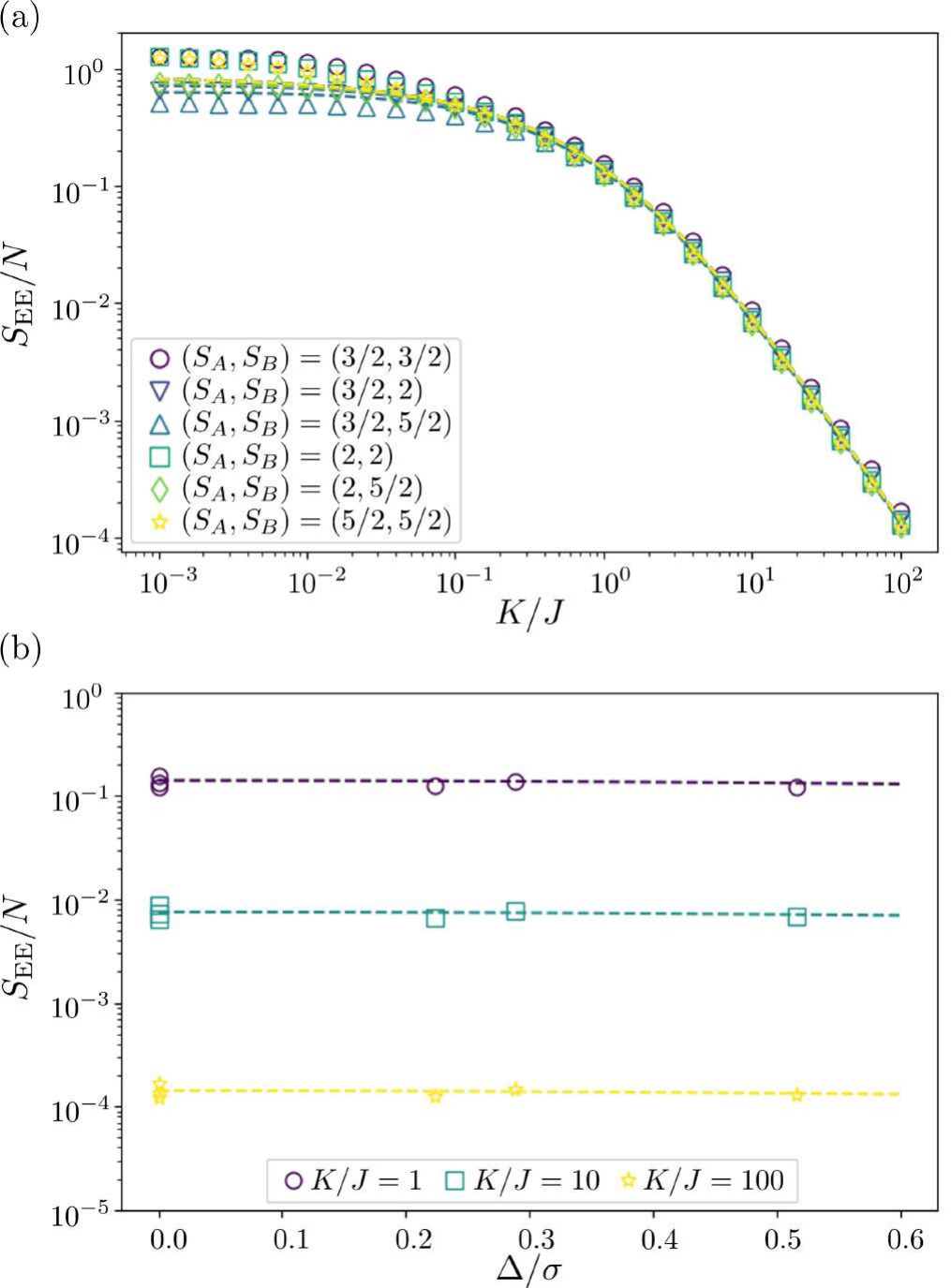}
	\caption{%
		\label{fig:3}
		Comparison of entanglement entropy between the analytic approximation [Eq.~(\ref{eq:entropy})] (dotted line) and DMRG simulation (marker) with $50$ sites considering six possible combinations of spins from $3/2$ to $5/2$. 
		For the DMRG simulation, the plotted values are the entanglement entropy obtained between left and right parts of the spin chain divided by the center-cut.
		(a) Plot of $S_{\textrm{EE}}$ as a function of the ratio $K/J$ for each combination of spins.
		(b) Plot of $S_{\textrm{EE}}$ as a function of the ratio $\Delta/\sigma = (S_B - S_A) / \sqrt{S_A S_B}$, for $K/J = 1, 10, 100$.
		The dotted line decreases, albeit very slowly, as $\Delta/\sigma$ increases.
	}
\end{figure}

Figure~\ref{fig:3} shows plots of the entanglement entropy per unit cell for ferrimagnet spin chains with respect to $K/J$ and $\Delta/\sigma$, based on the analytic relation~\eqref{eq:entropy} and DMRG simulation.
For the analytic results, the density of the entanglement entropy $S_\text{EE}/N$ [Eq.~\eqref{eq:entropy}] is used, whereas, for the DMRG results, the entanglement entropy between the left and the right parts of the spin chain divided by the center-cut is used. In Fig.~\ref{fig:3}(a), the entanglement entropy $S_{\textrm{EE}}$ of ground states from DMRG simulations decreases as $K/J$ increases.
The dotted lines fit with the corresponding numerical data for large anisotropy ($K/J>1$), whereas the values differ in the small anisotropy region ($K/J<1$).
For $K/J<1$, the analytic relation underestimates $S_{\textrm{EE}}$ in antiferromagnetic situations, but it overestimates in ferrimagnetic cases.
Hence the magnon description provides a better approximation of the ground state for large $K/J$.
In Fig.~\ref{fig:3}(b), $S_{\textrm{EE}}/N$ is plotted against the ratio $\Delta/\sigma$ in the large anisotropy region.
The analytic expression approximates $S_{\textrm{EE}}/N$ precisely, as the numerical result for each combination of spins $(S_A, S_B)$ follows the dotted lines.

\section{Discussion}
\label{sec:5}

We have investigated the entanglement entropy between the sublattices in the ground state of the ferrimagnet spin chain with uniaxial anisotropy.
The analytic equations for the ground state energy and entanglement entropy depend on the spin difference per geometric mean of spins ($\Delta/\sigma$) and the anisotropy ($K/J$).
They are formulated through the thermodynamic limit on the calculations from diagonalizing Hamiltonian in squeezed magnon representation.
The analytical results are verified numerically by using DMRG, which approximates the entanglement with high precision.

To understand the ferrimagnetic nature in the entanglement entropy per unit cell $S_{\textrm{EE}}/N$, it is useful to focus on the analytic relation.
Recall that $S_{\textrm{EE}}/N$ only depends on a parameter $0 < m \le 1$ which is the inverse of $(1+K/J)\sqrt{1+(\Delta/2\sigma)^2}$.
One can show that $S_{\textrm{EE}}/N$ is a decreasing function of $m$ since the thermodynamic limit of the R\'enyi entropy $S_{\alpha} = \ln\Tr(\rho_A^\alpha)/(1-\alpha)$ per unit cell is a decreasing function of $m$ for a fixed $\alpha$ and $S_{\textrm{EE}} = \lim_{\alpha\to1} S_{\alpha}$~\cite{renyi_measures_1961}.
From this, $S_{\textrm{EE}}/N$ tends to decrease if $K/J$ or $\Delta/\sigma$ increases.
The influence of the anisotropy against $S_{\textrm{EE}}$ is explicitly shown in Fig.~\ref{fig:3}(a).
In contrast, a small change of $\Delta/\sigma$ rarely changes the amount of $S_{\textrm{EE}}$ in Fig.~\ref{fig:3}(b).
Hence one can deduce that $S_{\textrm{EE}}$ is stable under a small variation of $\Delta/\sigma$ in large anisotropy region ($K/J>1$).

To consider the ferrimagnet with large $\Delta/\sigma$, one can increase $S_B$ while fixing $S_A$.
Then the ground state configuration resembles a ferromagnetic configuration on sublattice $B$, and the intersublattice $S_{\textrm{EE}}$ decreases to zero since $\Delta/\sigma$ diverges.
This implies that the sublattices $A$ and $B$ become separable as the spin difference is large enough.
Conversely, when there is no spin difference, the analytic equation is reduced to that of an ordered antiferromagnet, as in Ref.~\cite{hartmann_intersublattice_2021}.
Hence one can deduce that the entanglement entropy of a ferrimagnet is between that of a ferromagnet and an antiferromagnet.

It can be observed that the entanglement entropy of the ground state of the ferrimagnet spin chain satisfies an area law, as expected in various short range systems~\cite{eisert_colloquium_2010}.
The area law of $S_{\textrm{EE}}$ for a subsystem $I$ for a quantum state in 1D system $S$ can be stated as $S_{\textrm{EE}} = O(|\partial I|)$ where $\partial I$ is the subset of $I$ that interacts with $S \backslash I$.
In our ferrimagnet spin chain system, the boundary sizes are $N$ and $1$ under the consideration of the intersublattice (analytic relation) and center cut (DMRG), respectively.
From Fig.~\ref{fig:3}(a), it is evident that $S_{\textrm{EE}}$ of the ground state of the ferrimagnet spin chain follows area law scaling.

We conclude the paper with two open questions.
First, in the ferrimagnetic spin chain with small anisotropy ($K/J<1$), the analytic result for the entanglement entropy deviates from the numerical value.
A quantitative understanding of the difference can be pursued in future research.
In addition, the method of Ref.~\cite{hartmann_intersublattice_2021} used in this work to obtain the intersublattice entanglement entropy of ferrimagnets can be extended to other spin chain models.
A potential direction for future research could involve a system with Dzyaloshinskii–Moriya interaction term to investigate the entanglement structure of non-collinear spin states.

\begin{acknowledgments}
	This work was supported by Brain Pool Plus Program through the National Research Foundation of Korea funded by the Ministry of Science and ICT (NRF-2020H1D3A2A03099291), by the National Research Foundation of Korea (NRF) grant funded by the Korea government (MSIT) (NRF-2021R1C1C1006273), and by the National Research Foundation of Korea funded by the Korea Government via the SRC Center for Quantum Coherence in Condensed Matter (NRF-RS-2023-00207732).
\end{acknowledgments}

\bibliography{references}

\end{document}